\documentclass[reprint]{revtex4-1}
\usepackage{graphicx}
\usepackage{dcolumn}
\usepackage{bm}
\usepackage{amsmath}
\usepackage{amssymb}
\usepackage[utf8]{inputenc}

\begin{document}

\title{Physical aspects of symmetry breaking in Bose gases at thermal equilibrium}

\author{Alexej~Schelle}
\affiliation{Senior Lecturer @ IU Internationale Hochschule, Juri-Gagarin-Ring 152, D-99084 Erfurt}

\date{\today}% It is always \today, today,

\begin{abstract}

The theory of non-interacting Bose gases is supplemented by a numerical quantum field description with a two-dimensional non-local order parameter that allows the modeling of wave-like atomic correlations and interference effects in the limit of low atomic densities.  
From the present model, it is possible to explain symmetry aspects of non-interacting and very weakly interacting Bose gases in the limit of fluctuating particle numbers, like the forward propagation of time and the relation to the breaking and preservation of phase gauge symmetry in solids.
In the present formalism, the propagation of one-directional time arises from the pre-defined and equivalent convergence of independent quantum fields towards the Boltzmann equilibrium, and it is shown that Glauber coherent states are related to the definition of the quantized field. 
Coherently coupling condensate and non-condensate parts as a direct consequence of the increasing quantum coherence time between the different quantum field components in the Bose gas from cooling to below the critical temperature, the present model describes symmetry breaking, which is originally known from the definition of a specific gauge field 
from Elitzur's theorem for local gauge fields, as a global physical rather than a purely formal mathematical process.
 
\begin{description}
\item[Purpose] Preprint version of the article
\end{description}
\end{abstract}

\maketitle

\section{Introduction}

The theory of non-interacting Bose gases shows a long history of modeling aggregate states and phase transitions of many-particle quantum systems confined to external trapping potentials such as photonic Bose-Einstein condensates \cite{ref-1, ref-2}. 
For system parameters that assume a constant number of particles in the Bose gas, results and concepts of ideal gas theory in the canonical ensemble have, in particular, been extensively applied and reformulated to address fundamental questions of condensed matter theory with particles of definite masses, 
such as the condensation of atoms to a Bose-Einstein condensate below the critical temperature \cite{ref-3}. 
Though the assumptions of kinetic ideal gas modeling make it possible to understand the characteristics and dynamics of a very weakly interacting many-particle atomic quantum system,
finding a parameter regime of low enough densities and large temperatures that still shows condensation of atoms into a macroscopic ground state population of the typically energetically lowest-lying single-particle energy mode in the particle spectrum 
in a real experiment often makes it challenging to produce reliable quantitative results for interacting systems when only calculating numerical results for non-interacting particles \cite{ref-4, ref-5}.

Theoretically, the theory of ideal gases in the canonical ensemble provides a straightforward tool to understand and model the dynamics of many particles for the entire range below and above the critical temperature for Bose-Einstein condensation \cite{ref-6, ref-7}.    
Such experimental realizations of Bose-Einstein condensates with minimal interaction strength (described within first-order perturbation theory by the s-wave scattering length) are typically well-covered by the model results of kinetic gas theories, i.e.
in the limit of low densities and high temperatures at zero interaction strength \cite{ref-8}. 
Yet the fundamental results and applicational practices of ideal gas theory are well established, so far, more needs to be known about simple theoretical formulations to explain the relation of reversible 
microscopic many-particle dynamics and the time direction of many-body quantum systems \cite{ref-9, ref-10}. 

In the context of non-interacting gases, irreversibility is understood as entropy maximization at the Boltzmann equilibrium, as formulated by the second law of thermodynamics.    
A daily observation of experiments is that time passes irreversibly in one direction, which can e. g. be counted with an atomic clock that measures the pings of oscillating photons or atoms, respectively, at the edges of a microwave cavity or by the number of decays of atoms in an atomic crystal, to define a unit of time.
However, since non-interacting gases do not obey dissipation or inelastic collisions, the dynamics of ideal many-body processes are completely reversible in time, and the fundamental process for breaking time or phase gauge symmetries must not depend on the interactions of a many-particle system \cite{ref-11}. 

For that purpose, however, the theory of ideal gases that assumes a constant number of particles in the Bose gas has to be extended to the case of lowly fluctuating particle numbers.
In this limit, it is in particular possible to combine the theory of ideal gases with the concept of forward propagating time, which is a highly non-trivial conceptual task, since, as argued before, time principally obeys a (reflection) symmetry for non-interacting atomic systems.
Relating the theory of non-interacting Bose gases to the concept of spontaneous symmetry breaking in Bose-Einstein condensates in terms of breaking the phase gauge invariance in the quantum system \cite{ref-12} requires the assumption of a fluctuating total number of particles 
in different realizations of the physical setup in a temperature limit below the critical temperature, where the canonical and the grand-canonical ensembles are no longer equivalent. 
To supplement physical aspects of spontaneously broken gauge symmetries in non-interacting Boses gases that are currently not straightforwardly explained by the standard theory of ideal gases, the present work explains the context of the fundamental results of a number-conserving quantum field theory to representations of Bose-Einstein condensates in terms of Glauber coherent states in the limit, where the particle number is finite and allowed to fluctuate slightly around the mean value defined by a macrcoscopic average value.

In the sequel to the present theory, first, the symmetry-preserving unitary time evolution according to the theory of non-interacting Bose gases is introduced, before extensions to the description of non-interacting Bose gases in terms of a non-local order parameter in complex number representation are explained.
In the framework of this non-local quantum field theory for a total average number of $N$ particles at a definite temperature $T$, decay, and oscillation rates associated with partial chemical potentials can be numerically derived in complex number representation. 
This quantum model, in particular, explains the one-directional propagation of time with an average zero phase of the quantum wave field from the interference of ideally disjoint forward and backward propagating wave fields at the Boltzmann equilibrium \cite{ref-9} with vanishing total chemical potential.
The scaling of the coherence time indicates that the total quantum field is composed of a joint combination of two field components (a field generated from the condensate mode population and a second field created from the rest of the single particle modes), leading to a symmetric total field distribution 
that builds an extended ring in the complex plane above the critical temperature. 
This field decays into a point-like field of zero average value as a consequence of the increasing coherence time between condensate and non-condensate subspaces below the critical temperature, when the coherence time approaches values of the minimal time uncertainty. 

The presented theory allows the numerical calculation for realizations of the quantum field that defines Glauber coherent states for a non-interacting Bose gas below as well as above the critical temperature for Bose-Einstein condensation \cite{ref-13}.
Thereby, it is shown ab initio that the phase of the distribution of coherent field states is distributed around zero phases at the Boltzmann equilibrium with a constant total number of $N$ particles at a definite temperature $T$ and vanishing total chemical potential $\mu$, 
while phase-coherent states with arbitrary phase can, in principle, be realized from Monte-Carlo simulations of the quantum field that follows equilibrium states with lower total entropy corresponding to non-zero values of the chemical potential $\mu\ne0$.

Finally, this explains that forward time propagation and spontaneously broken gauge symmetries of the Schr\"{o}dinger equation's gauge fields occur as a physical process below the critical temperature, originating from quantum coherence of the quantum states in a theory with non-local order parameters 
that gauges the average total field to zero and further implements the physical boundary condition of a Boltzmann thermal equilibrium into the theoretical model. 
The model applies in the limit of non-interacting Bose gases with also lowly fluctuating particle numbers, i.e. nearly ideal Bose gases below the critical temperature. 

\section{Theoretical Model}

In the following, we consider a bosonic gas of $N$ non-interacting atoms confined in an external harmonic trapping potential. 
A supplementary model for the ideal gas theory is presented that covers the description of the non-reversible forward-time evolution of a non-interacting many-particle system below and above the critical temperature for Bose-Einstein condensation in the limit, 
where the total particle number is allowed to fluctuate from one realization to the next.

The model describes the quantum field of a non-interacting Bose gas in terms of a two-dimensional vector that includes either one order parameter for the quantum system's forward and backward propagating unitary time evolution.
Applying this mathematical structure, the interference of the two field components defines the forward propagation of time provided that the system is considered to follow the Boltzmann equilibrium.
Time is measured in quantized physical units of seconds, which means that the extended ideal gas model can be interpreted as a model description for elementary quantum clocks. 
Modeling the quantum field at finite temperature, assuming thermal equilibrium, leads to numerical realizations of Glauber coherent states with an initial phase defined by the (pre-defined) zero phase of the quantum field.     

\subsection{Unitary time evolution}

In the standard theory of quantum mechanics using the concept of second quantization \cite{ref-15}, the time evolution of the system's quantum state is mathematically described by
the unitary operator

\begin{equation}
\label{eq.1}
\hat{U}(t) = {\rm exp}\left [-\frac{i\hat{\mathcal{H}}t}{\hbar}\right] \ , 
\end{equation}
where $\hat{\mathcal{H}} = \sum_{\epsilon_{\bf k}}\epsilon_{\bf k}\hat{a}^\dagger_{\bf k}\hat{a}_{\bf k}$ represents the energy operator of the quantum many-particle system in second quantization.
In Eq. (\ref{eq.1}), the variables $\epsilon_{\bf k}$ define the energy of a single particle in the external trapping potential, and the operators $\hat{a}^\dagger_{\bf k}$ and $\hat{a}_{\bf k}$ describe the creation and destruction 
of quantum states of a single particle that populates a certain state with the corresponding energy $\epsilon_{\bf k}$. 
From the eigenvalue equation

\begin{equation}
\label{eq.2}
\hat{H} \vert \Psi_{\bf k}\rangle = \epsilon_{\bf k} \vert \Psi_{\bf k}\rangle \
\end{equation}
one finds that the single-particle quantum states are defined as the eigenstates of the Hamilton operator $\hat{H}$ of a single particle in the weakly interacting Bose gas.
We further assert that the single-particle spectrum is approximately given by the eigenenergies of the harmonic oscillator

\begin{equation}
\label{eq.3}
\epsilon_{\bf k} = \left(k_x\omega_{x} + k_y\omega_{y} + k_z\omega_{z} + \frac{3}{2}\right) \ ,
\end{equation}
with $\omega_{x}$, $\omega_{y}$, and $\omega_{z}$, the trapping frequencies of the external confinement, for very weakly interacting Bose gases. 
Indeed, given Eqs. (\ref{eq.1}) - (\ref{eq.3}), the reversible time evolution of the quantum system can be well described by the Schr\"{o}dinger equation that describes the time evolution of 
a many-body quantum state $\vert \psi_{\bf{k}} \rangle = \vert\psi_{\bf{k}_1}\rangle\otimes ... \vert\psi_{\bf{k}_N}\rangle$. 

\subsection{Quantum field and order parameter}

The aim of the present quantum mechanical description for a Bose gas in second quantization is to find ideally only one single physical quantity that describes the condensation characteristics of the quantum system.
To this end, from Eq.~(\ref{eq.3}), one can define a general quantum state

\begin{equation}
\label{eq.4}
\vert \Psi \rangle = \sum_{\bf k}b_{\bf k}\vert \Psi_{\bf k}\rangle \ 
\end{equation}
that describes realizations of the many-body quantum states of $N$ non-interacting particles with random coefficients $b_{\bf k}$.
From Elitzur's theorem, \cite{ref-16}, one may learn that there exist no spontaneously broken local gauge symmetries without fixing the field gauge, i.e., since the phase gauge symmetry of three-dimensional quantum systems is suspected to be broken at the phase transition for Bose-Einstein condensation,
it is formally motivated to define a non-local spatially averaged order parameter for the Bose gas that describes phase transitions and the corresponding symmetry breaking of the gauge fields globally with a single numerical value \cite{ref-17} .

Projecting Eq.~(\ref{eq.4}) onto the basis $\vert \bf{r} \rangle$, and integrating over three-dimensional space (spatial average), from the coefficients $b_{\bf k} = {\rm exp}\left[ -\frac{i(\epsilon_{\bf k}-\mu_{\bf k})t}{\hbar}\right]$, one 
obtains the following definition for an order parameter,

\begin{equation}
\label{eq.5}
\psi = \sum_{\epsilon_{\bf k}}c_{\bf k~}{\rm exp}\left[\frac{-i\mu_{\bf k}t}{\hbar}\right] \ ,
\end{equation}
where new (normalized) random coefficients $c_{\bf k} = 0...1$ are introduced.
In Eq. (\ref{eq.5}), the parameters $\mu_{\bf k}$ define the inverse chemical potentials of the atoms in the Bose gas.
Up to this formal derivation, the model can, in principle, be applied to any quantum system with eigenstates that fulfill the Schr\"{o}dinger equation. 

In a gas of $N$ non-interacting atoms, there is no ad hoc theoretical constraint that defines a clear direction of time. The system passes a symmetric unitary time evolution according to the Schr\"{o}dinger equation.
Therefore, the order parameter may be defined for both directions in time, so that the vector

\begin{equation}
\label{eq.6}
\vec{\Psi}(t) = \left(\begin{array}{c} \psi_+ (t) \\ \psi_- (t) \end{array}\right) \ ,
\end{equation}
describes an extended point of view for the order parameter, i. e. an order parameter $\psi_+(t) = \psi (t)$ that describes the forward and $\psi_-(t) = \psi (-t)$ the backward time evolution of the average quantum field.
At the thermal Boltzmann equilibrium, which is formally defined by the variable transformation $it = \beta\hbar = -it$ in the formal limit, where $\hbar \rightarrow 0$, the two components of the order parameter effectively equal, and the order parameter becomes

\begin{equation}
\label{eq.7}
\psi = \psi_+ = \psi_- = \frac{1}{2}(\psi_+ + \psi_-) = \sum_{\epsilon_{\bf k}}c_{\bf k~}{\rm exp}\left[-\beta\mu_{\bf k}\right] \ ,
\end{equation}
where $\beta = (k_BT)^{-1}$ is related to the inverse thermal energy of the Bose gas. 
Hence, in the limit where $\mu = 0$, the (phases of the) forward and backward propagating partial wave fields of the order parameter are equal and no longer disjoint, and in particular, thermal mixing of the two components occurs.

\subsection{Elementary model for time evolution}

An elementary understanding of the time evolution for the Bose gas that is mathematically described in terms of a non-local order parameter can be obtained by considering the Bose gas as a very simple model for an elementary quantum clock. 
From forward and backward propagating photons in a microwave cavity, measurements of the number of outcoupled photons at the edges of the microwave cavity after a certain (oscillation) time interval $\tau_0$ 
from the interaction with the external environment can be obtained. 
At each measurement cycle, the photon field can hence be assumed to get in thermal equilibrium with a second source from the contact with the external environment, realized by the emission of photons that are outcoupled from the microwave cavity after each oscillation cycle.    
In such a setup, both the forward as well as backward propagating fields define a (forward and backward) direction of time, and the symmetry-preserving time variable evolves in units of a quantized variable $\tau_0$. 

It is only at the Boltzmann equilibrium with vanishing chemical potential (which is derived for the equilibrium with the thermal external environment from particle number conservation at a definite temperature), where both fields are exactly equal at 
multiples of this time interval, and the total system interacts with the thermal environment. 
This can be regarded as a measurement process that defines a macroscopic variable with the temperature $T$. 
Thus, the two field components $\psi_+(t)$ and $\psi_-(t)$ entail a symmetry property concerning a period $\tau_0$ that leads to the (parameter-dependent) quantization of the time variable, $t=m\tau_0$, with integer values of $m$, where 

\begin{equation}
\label{eq.8}
\psi_+(t=m\tau_0) = \psi_-(t=m\tau_0) \ .
\end{equation}
Coherent phases $\phi_{\bf{k}}(t)$ in Eq. (\ref{eq.5}) can be directly mathematically obtained from the partial chemical potentials $\mu_{\bf{k}}$ of the atomic quantum field modes, through the relation $\phi_{\bf{k}}(t) = {\rm Re}\lbrace\frac{\mu_{\bf{k}}t}{\hbar}\rbrace$. 
Hence, the partial chemical potentials can be quantified by the conservation of the average total number of particles in the Bose gas, from the fact that

\begin{equation}
\label{eq.9}
\sum^{\infty}_{j\ne0} z^{j}(\mu_{\bf k}) \left[\prod_{l=x,y,z} \frac{1}{1-{\rm e}^{-j\beta\hbar\omega_l}} - 1\right] = (N-N_0) ,
\end{equation}
with $z(\mu_{\bf k})= {\rm e}^{\beta\mu_{\bf k}}$ the fugacity of the particles and $\omega_{\bf k}$ the photon frequency of the particles in the Bose gas (in mode direction $l = x, y, z$). 
This finally leads to numerical estimates of the coherence time between forward and backward propagating field components that indicate the range of a few hundred microseconds in the present setup. 

\subsection{Representations with Glauber coherent states}

\begin{figure}[t]
\begin{center}
\includegraphics[width = 8.00cm, height = 6.00cm, angle=0.0]{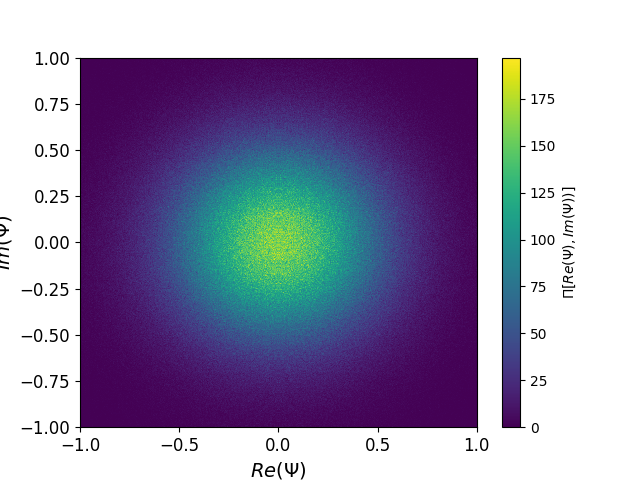} 
\caption{(Color online) Monte-Carlo simulation of the distribution $P(\alpha)$ with $5\times10^6$ samples of the (combined) wave fields as defined in Eq. (\ref{eq.7}) to numerically model the quantum field distribution in units of Glauber coherent quantum states.
Parameters for the numerical simulation are set to $\omega_x = 2\pi\times125.0$ Hz, $\omega_y = 2\pi\times75.0$ Hz and $\omega_z = 2\pi\times25.0$ Hz at $T=15.0$ nK with $N = 5000$ atoms and few hundred considered eigenmodes.}
\label{figzero}
\end{center}
\end{figure}

A very intuitive and well-established representation of photonic eigenstates of the annihilation operator is defined by the Glauber coherent quantum states

\begin{equation}
\label{eq.10}
\vert \alpha \rangle =  {\rm e}^{-\vert\alpha\vert^2}\sum_n\frac{\alpha^n}{\sqrt{n!}}\vert n\rangle \ ,
\end{equation}
which in the present formalism can be related to the quantum field of the non-interacting Bose gas at thermal equilibrium with the representation of the field in the complex number plane $\alpha = \sqrt{N}\psi = \sqrt{N}\vert\psi\vert{\rm e}^{i\phi}$ and 
the distribution $P(\alpha)$ which leads to 

\begin{equation}
\label{eq.10a}
\hat{\rho} =  \int {\rm d}^2\alpha~P(\alpha)\vert\alpha\rangle\langle \alpha \vert \ ,
\end{equation}
where the variable $\phi$ defines a priori random global phase \cite{ref-18}.
A numerical representation of the so-defined distribution $P(\alpha)$ in terms of Glauber coherent states is shown in Fig.~\ref{figzero}.

From the Schr\"{o}dinger equation, which defines the time evolution in terms of the equation

\begin{equation}
\label{eq.11}
i \hbar \frac{\partial \vert \psi \rangle}{\partial t} =  \hat{\mathcal{H}}\vert \psi \rangle \ ,
\end{equation}
one may deduce that the formal solution

\begin{equation}
\label{eq.12}
\vert \psi \rangle = {\rm exp}\left[\frac{-i\hat{\mathcal{H}}t}{\hbar}\right] \vert \psi_0 \rangle
\end{equation}
defines a stationary state of the Bose gas, given the initial condition $\vert \psi_0\rangle$.

\begin{figure}[t]
\begin{center}
\includegraphics[width = 4.25cm, height = 3.50cm, angle=0.0]{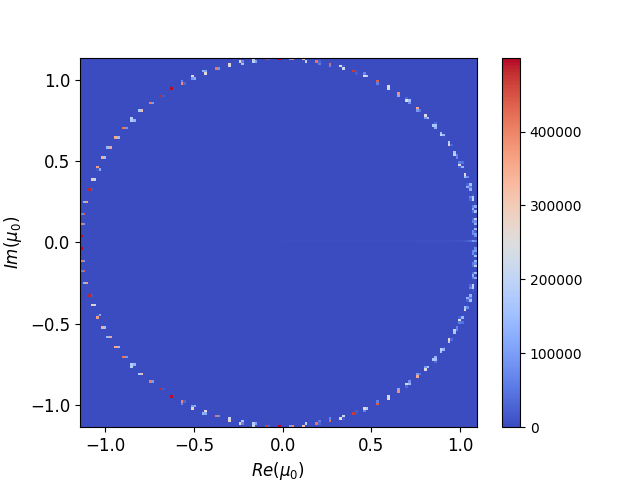} 
\includegraphics[width = 4.25cm, height = 3.50cm, angle=0.0]{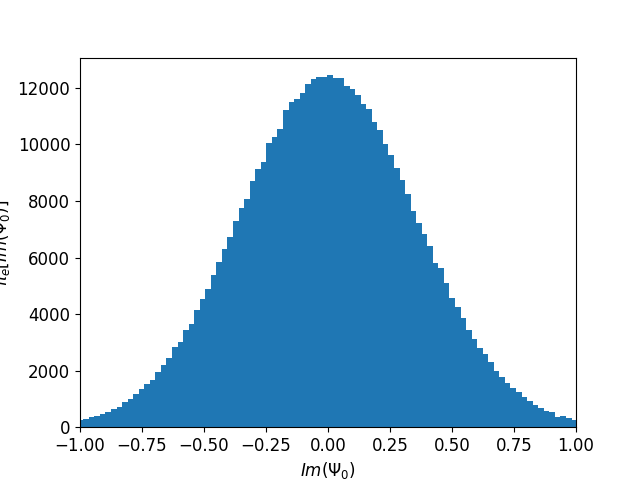} \\
\includegraphics[width = 4.25cm, height = 3.50cm, angle=0.0]{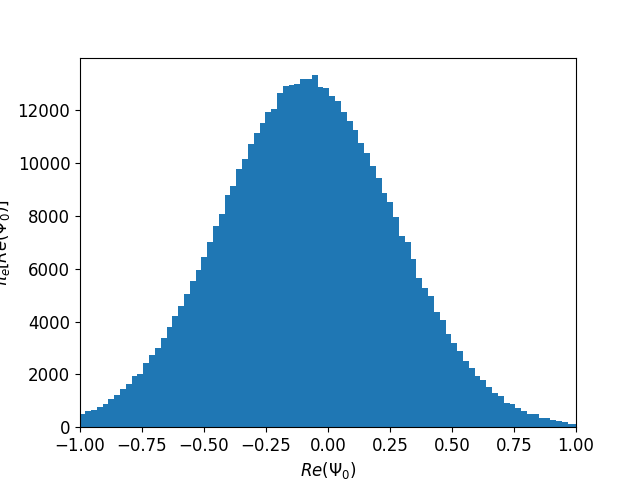} 
\includegraphics[width = 4.25cm, height = 3.50cm, angle=0.0]{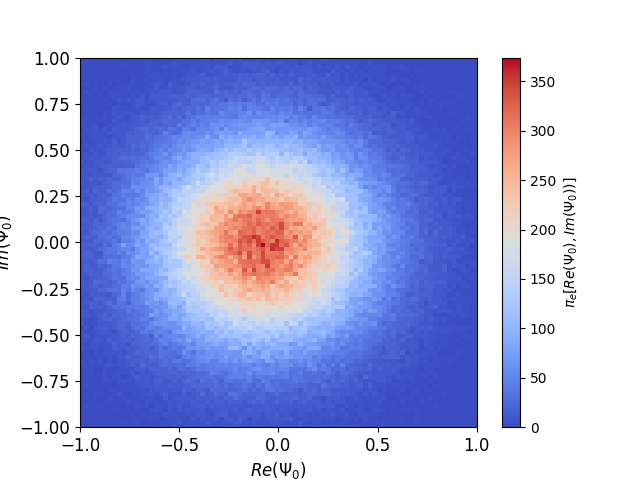} 
\caption{(Color online) Figures highlight the broken phase gauge symmetry in the complex plane. 
The upper left figure shows the distribution of the condensate chemical potential at thermal equilibrium in the complex plane, which is composed of a ring with a constant radius in the complex plane and a symmetry-breaking constant part in the direction of the real axis. 
Upper right and lower left figures show the distribution of the imaginary parts (upper right) and the real parts (lower left) of the condensate wave field obtained from Eqs. (\ref{eq.7}), (\ref{eq.9}) und (\ref{eq.16}).
The lower right figure explains the broken symmetry of the wave field's aggregate phases, i.e., condensate and non-condensate parts (left-shifted and right-shifted circles in the representation of numbers in the complex plane). 
Parameters for the numerical simulation are set to $\omega_x = 2\pi\times125.0$ Hz, $\omega_y = 2\pi\times75.0$ Hz and $\omega_z = 2\pi\times25.0$ Hz at $T=25.0$ nK with $N =5000$ atoms.}
\label{figone}
\end{center}
\end{figure} 

\subsection{Correlations of the relative phase}

As illustrated in terms of an analytical formulation in the previous chapter, correlations between phases of the forward and backward propagating components 
of the non-local order parameter gives rise to relationships between irreversible macroscopic quantities, such as temperature or time, with the 
microscopic description of the Bose gas.
Numerically, phase correlations between the two different components $\psi_+$ and $\psi_-$ can be calculated ab initio from the equation

\begin{equation}
\label{eq.13}
c(\chi^\star, \chi) = \chi^*(\Delta\phi, \beta)\chi(\Delta\phi, \beta) \ ,
\end{equation}
with the parameter $\chi(\Delta\phi, \beta)$ defined as the coherent sum of forward and backward propagating quantum fields $\chi(\Delta\phi, \beta) = \psi_{+}(\beta) + {\rm e}^{i\Delta\phi}\psi_{-}(\beta)$.
An approximate analytical formula for the correlation function $c(\chi^\star, \chi)$ is given by a Gaussian distribution around the average zero phases of $\chi(\Delta\overline{\phi} = 0, \beta)$, i.e.

\begin{equation}
\label{eq.14}
c(\chi^\star, \chi) = c(\Delta\phi) \approx \frac{{\rm exp}\left[-\Delta\phi^2 \right]}{\mathcal{N}} \ ,
\end{equation}
where $\mathcal{N}$ is a normalization constant.

More sophisticated numerical analysis shows mathematically more complex correlation functions between forward and backward propagating wave fields with zero average phases, 
which is shown in the quantitative analysis of the next chapter III.
The Schr\"{o}dinger equation in Eq. (\ref{eq.11}) once more indicates that spontaneous symmetry breaking isn't related to the occurrence of a random, but to a pre-defined phase of the quantum field, 
since the physical constraints in a Bose gas intrinsically pick out a specific zero phase due to quantum coherence at the Boltzmann equilibrium below the critical temperature, 
while the total distribution of all possible gauge fields is still symmetry-preserving.  

\subsection{Spontaneous phase gauge symmetry breaking}

The concept of spontaneous phase gauge symmetry-breaking explains a certain sudden change in the symmetry properties of the gauge fields that describe 
a solid below the critical temperature.
As an example, one may consider a magnet that changes its magnetization from an average zero to a well-defined direction with components of spin up and spin down 
in the z-direction of the magnet below a specific Curie temperature for a characteristic phase transition.
While the experimental observation of symmetry breaking is understood straightforwardly, the formulation of a consistent theory that explains spontaneous symmetry breaking as a physical rather than a purely mathematical process is highly non-trivial.

\begin{figure}[t]
\begin{center}
\includegraphics[width = 8.5cm, height = 7.0cm, angle = 0.0]{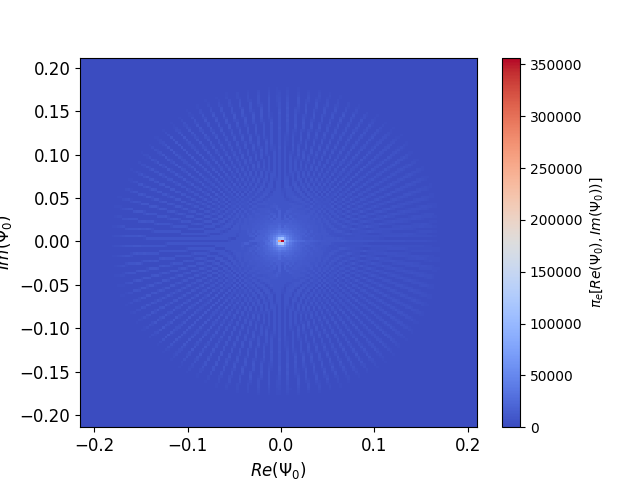} 
\caption{(Color online) Figure shows the distribution of the (unweighted) condensate wave field modes with the constraint of the Boltzmann equilibrium at the same experimental conditions as in Fig. \ref{figone}.
Instead of the continuously distributed wave field in the complex plane, the wave field shows a star-shaped linear spreading from zero to a finite value in the direction of quantized angles $\phi$.}
\label{figtwo}
\end{center}
\end{figure} 

Any initial state of the formal solution in Eq. (\ref{eq.12}) with a specific initial phase defined by $\psi_0 = \vert\psi\vert{\rm e}^{i\phi_0}$ defines a separate and non-unique solution of the Schr\"{o}dinger equation.
The corresponding gauge fields are symmetric as long as no physical constraint is assumed for the quantum field's chemical potential and are defined by Glauber coherent states that all preserve the symmetry of the Schr\"{o}dinger equation.
Considering only manifolds of the Boltzmann equilibrium with vanishing chemical potential $\mu = 0$ leads to projections onto subspaces of the gauge fields as a consequence of a physical constraint, the Boltzmann equilibrium, i.e., to the distribution $P(\alpha) = \delta(\alpha-\alpha_0)$, without the requirement of 
formally defining a certain symmetry-breaking gauge for the wave field.
Please note, however, that this does not define a symmetry-breaking process of the quantum system, since the symmetry of the total quantum field is preserved.
The projection of the total wave field onto the manifold of the Boltzmann equilibrium finally implies the propagation of a one-directional time, i. e. "the forward direction of time".
From Eqs. (\ref{eq.6}) - (\ref{eq.8}), it is thus likely to understand that physical time passes in only one direction, since at the Boltzmann equilibrium, where disjoint forward and backward propagating components 
of the quantum field are effectively mixed and equal, the partial waves interfere and lead to a well-defined mathematical form of the quantum field that doesn't diverge numerically, also below the critical temperature for Bose-Einstein condensation.

\begin{figure}[b]
\begin{center}
\includegraphics[width = 8.0cm, height = 5.5cm, angle=0.0]{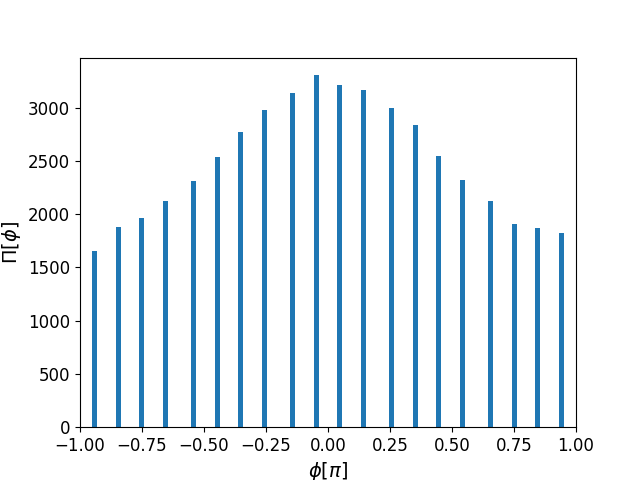} 
\caption{(Color online) Shown is the symmetric frequency comb spectrum of the total wave field that occurs from the constructive interference of partial matter waves that lead to one-directional time propagation according to the theoretical model.
The periodicity of the directional propagating wave field can be derived analytically, as expressed in Eq. (\ref{eq.17}).}
\label{figthree}
\end{center}
\end{figure} 

The breaking of the \textit{relative} phase gauge symmetry between condensate and non-condensate fields during the phase transition can be theoretically understood as an effect of quantum coherence below the critical temperature that changes the total field from a state of the form $\psi = \psi_0 \cup \psi_\perp$ to a coherent superposition 
of the components $\psi_0$ and $\psi_\perp$.
Considering a quantum field without local representations of the particle's wave functions facilitates the formulation of the relative symmetry breaking between the different components (aggregate phases) of the quantum system. 
As will be illustrated numerically ab initio in the following chapter III, in mathematical terms, symmetry breaking in non-interacting Bose gases below the critical temperature can be understood by the introduction of a non-local order parameter as defined by a coherent sum of condensate and non-condensate fields, 

\begin{equation}
\label{eq.15}
\psi = \psi_0 + \psi_\perp \
\end{equation}
within the framework of the present number-conserving quantum field theory. 
Since a total number of $N$ non-interacting atoms is assumed, the total average field $\psi = 0$ (the gauge of the order parameter of the total field) vanishes (see e.g. Ref.~\cite{ref-12}).
This is, in particular, described and accounted for by Eq. (\ref{eq.9}) which represents the conservation of the average total number of particles.
From this assumption, one can recognize that the condensate and non-condensate aggregate phases of the Bose gas are separated by an angle of $\phi = \pi$, i.e.

\begin{equation}
\label{eq.16}
\psi_0 = - \psi_\perp \ ,
\end{equation}
hence, interpreting the spontaneously broken gauge symmetry of the order parameter below the critical temperature is possible in terms of the fields in Eq. (\ref{eq.16}).

\begin{figure}[t]
\begin{center}
\includegraphics[width = 8.0cm, height = 5.5cm, angle=0.0]{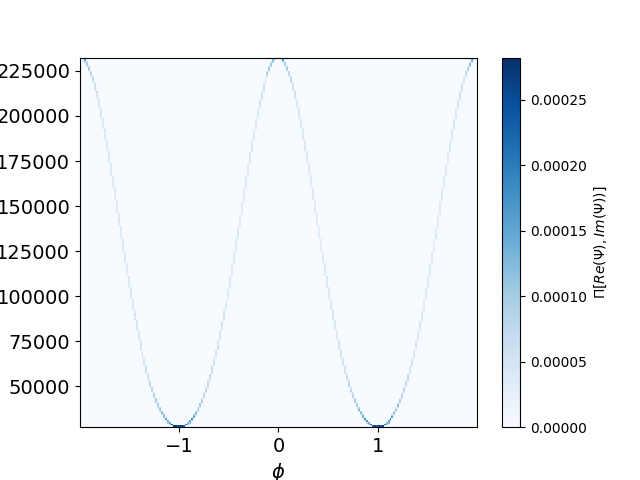} 
\caption{(Color online) Figure shows the phase correlations of the forward and backward field components of the wave field (in units of the circle number $\pi$) as defined in Eq. (\ref{eq.7}) from a numerical simulation.
As indicated by the numerical results, the phases of the two adjoint field components are correlated in quantized units of $2\pi$.
Different from phase correlations of two initially uncorrelated fields that are suddenly brought into contact, e.g., by a weak link such as in the setup of the Josephson model as described in Ref.~\cite{ref-7}, 
the distribution of the correlation function is not only predictable in terms of its maxima at $2\pi\times k$, but is also entirely deterministic and predictable in shape.}
\label{figfour}
\end{center}
\end{figure} 

\section{Numerical Results}

In this section, the phenomenon of broken gauge symmetry is analyzed in terms of correlations between adjoint forward and backward propagating wave fields.
In particular, the difference and interplay between broken phase gauge symmetry, phase correlations between locally distinct quantum fields, and the direction of time are illustrated and discussed.

\subsection{Broken aggregate phase gauge symmetry}

The concept of broken gauge symmetry in solids was originally introduced to understand and describe the occurrence of distinct phases between a critical temperature below which solids undergo a certain phase transition, as
characterized by an order parameter.
In the present theory, such phase transitions in atomic gases can be modeled by describing the transition of a mixed wave field $\psi = \psi_0 \cup \psi_\perp$ without coherences and a relative reference frame between condensate and non-condensate components of the Bose gas  
(that leads to a non-zero field distribution in the complex plane) to a coherently interfering condensate and non-condensate quantum field that is gauged to zero average (vanishing) values below the critical temperature for Bose-Einstein condensation.
In particular, below the critical temperature, the two sub-components, condensate and non-condensate, are coherently coupled within a certain coherence time (at each realization of the quantum field).  

Different from standard approaches known so far, within the present theory, it is, in particular, possible to explain that it is not the entire gauge field symmetry that is broken, but only the relative symmetry between condensate and non-condensate fields that are coherently interacting below the critical temperature.
In contrast, in the limit of large temperatures, the two components do not build a relative reference frame for symmetry breaking but are independently mixed, and therefore, the total quantum field builds a ball-shaped distribution in complex space.
For finite temperatures, there is theoretically no non-zero coherence; however, different field components only interfere destructively, if the coherence time is larger than the time uncertainty. 
This defines a physical transition for symmetry breaking when the coherence time exceeds this (minimal) uncertainty threshold.

Please note that this is the case only in three-dimensional setups, since in lower dimensions the single-particle wave functions cannot be assumed to obey a non-zero average.
Thus, for reduced dimensionality, the assumption that $\psi=0$ doesn't remain valid, i.e., one has to assume a general function $\psi = \psi(N-N_0)$.
The broken phase gauge symmetry of the (left-shifted) condensate quantum field is illustrated in the lower right section of Fig. \ref{figone}.
As compared to the star-shaped quantum field in Fig. \ref{figtwo}, it is visible that the (relative) symmetry is broken concerning the zero point axis, assuming thermal equilibrium. 

\begin{figure}[b]
\begin{center}
\includegraphics[width = 8.0cm, height = 5.5cm, angle=0.0]{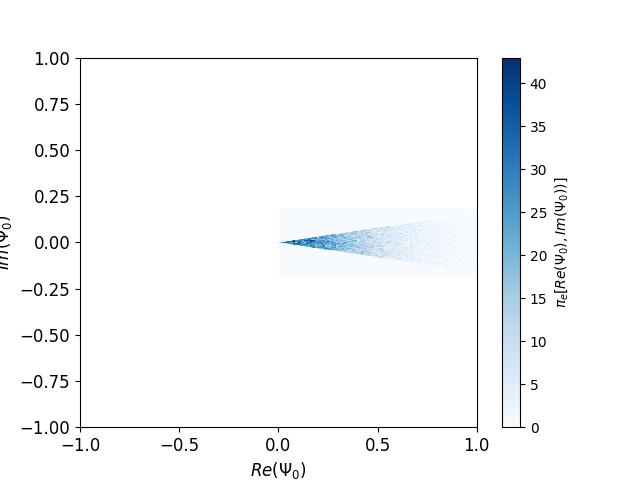} 
\caption{(Color online) Figure shows the condensate field in the direction of the positive real axis at the Boltzmann equilibrium with ideally 
${\rm Re}(\Psi_0) = 1.0$ and ${\rm Im}(\Psi_0) = 0.0$, i. e. an angle of $\phi_0 = -0.05\times\pi$ and $\phi_0 = +0.05\times\pi$ around the zero phase.
The so-obtained distribution indicates the field distribution that is relevant for the direction of time due to the distribution of the condensate wave field around the Boltzmann equilibrium.}
\label{figfive}
\end{center}
\end{figure} 

\subsection{Broken gauge symmetry from phase correlations}

As indicated from the derivation of Eq. (\ref{eq.8}), the time evolution of the quantum field is one-directional in units of the time scale $\tau_0$.
It is important to recall that the concept of one-directional time evolution can also be derived ab initio from the following numerical calculus.
Considering the correlation function in Eq. (\ref{eq.13}), one may straightforwardly verify that the initial phase of two (initially) uncorrelated quantum fields is distributed (quantized) around multiples of $2\pi$, see Fig. \ref{figfour}.
As discussed before in Ref.~\cite{ref-7}, the correlations around zero phase arise from the separate propagation of the independent fields towards the Boltzmann equilibrium, 
since at each calculation cycle, the two quantum fields are numerically calculated ab initio and independently.
The latter fact ensures that correlations around zero phase can not be only caused by the interference of pairwise independent particle matter waves, since the phases are not purely 
random, but pre-defined by the affinity of the wave fields to approach the Boltzmann equilibrium that leads to forward time propagation with similar distributions of partial phases (i. e. effectively zero relative phases).

Interference of partial waves at the Boltzmann equilibrium with one-directional time propagation, in particular, relates the frequency spectrum to the relative phase distribution of the two counter-propagating wave fields $\psi_+$ and $\psi_-$. 
From the fact that the partial phases follow the relation

\begin{equation}
\label{eq.17}
\frac{\phi_{\bf{k}}(\tau)}{\tau_0} = \omega m({\bf{k}}) + \omega_0 ,
\end{equation}\\
where $\omega m({\bf{k}})$ is the distribution of modes among the energy levels $\bf{k}$, one may conclude that the adjoint quantum fields interfere at multiples of the time scale $\tau_0$.
Please note that from the symmetry of the total wave field concerning the time scale $\tau_0$, i.e.. $\psi(t+\tau_0) = \psi(t)$, the spectrum of the spatially averaged quantum field (order parameter) of the theory remains preserved, as shown in Fig. \ref{figthree} for the same 
model parameters as in Fig. \ref{figone}.
Numerically, the manifold of the total wave field at the Boltzmann equilibrium with $\mu\rightarrow0^+$ can simply be extracted from the boundary condition that $\phi = \pm \epsilon \pi$, where the parameter $\epsilon$ ideally tends to zero and has been chosen $0.05$ in the numerical simulation 
shown in Fig. \ref{figfive}.

\subsection{Direction of time from constructive interference}

As we understand from the present model, the direction of time manifests itself as an interference process of forward and backward propagating wave fields $\psi_+$ and $\psi_-$.
Considering only quantum field states with complex number representations within the shown manifold, directional time is a propagating quantity that arises from the constructive interference of partial matter waves in the Bose gas.
The smallest unit time scales for the forward propagation of time are thus defined by the time scales at which partial matter waves interfere (from zero to a few microseconds), whereas, from destructive interference (dynamical localization in complex space), one may deduce that there exists 
a certain coherence time of the total field that defines a time scale for the decay of different realizations of the thermal Boltzmann equilibrium in the present parameter regime.

\begin{figure}[t]
\begin{center}
\includegraphics[width=6.5cm, height = 4.0cm, angle=0.0]{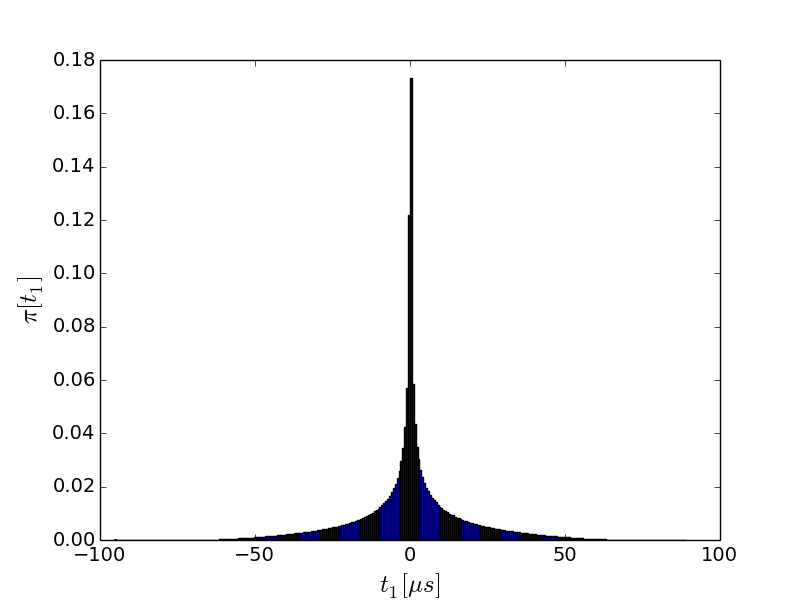} 
\includegraphics[width=6.5cm, height = 4.0cm, angle=0.0]{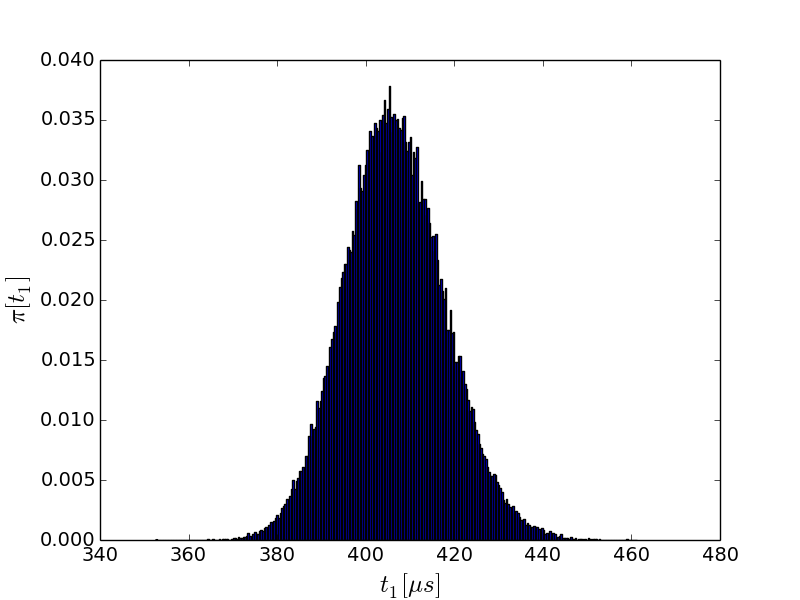} 
\caption{(Color online) Figure shows $10^5$ realizations of the imaginary parts (left figure - labeled $t_1 = \Gamma^{-1}_1$) and real parts (right figure - labeled $t_2 = \Gamma^{-1}_2$) 
of complex phase as defined in Eq. (\ref{coherence_time_2}) from random Markov sampling for a non-isotropic trap geometry $\omega_x,\omega_y = 2\pi\times42.0$ Hz and $\omega_z = 120.0$ Hz 
at temperature $T=5.0$ nK for $N = 1000$ atoms. 
The observation that $\Gamma_2t$ is non-zero at thermal equilibrium indicates the coupling of the equilibrium state to quantum states with $\Gamma_2t\ne0$ (non-classical correlations).}.
 \label{figsix}
\end{center}
\end{figure}
    
The expression $\mu t$ is not a real, but a complex number, that is, $\mu t\in\mathbb{C}$. 
Hence, in the framework of this quantum field theory as presented, the real part $\Gamma_1t$ describes the coherent phase evolution of the quantum particle at equilibrium (inverse oscillation frequency in the stationary state), corresponding to a ring of the fugacity spectrum, whereas $\Gamma_2 t$, the imaginary part (of $\mu t$) describes decay processes of the particle in the atomic cloud at or close to equilibrium (corresponds to a linear part of the fugacity spectrum), that is
 
\begin{equation}
\frac{\mu t}{\hbar\overline{\omega}} = \Gamma_1 t + i \Gamma_2 t ,
\label{coherence_time_2}
\end{equation}\\
where $\overline{\omega} = 0.5*(\omega_x\omega_y\omega_z)^{1/3}$. 

\begin{figure}[b]
\begin{center}
\includegraphics[width = 7.0cm, height = 4.5cm, angle=0.0]{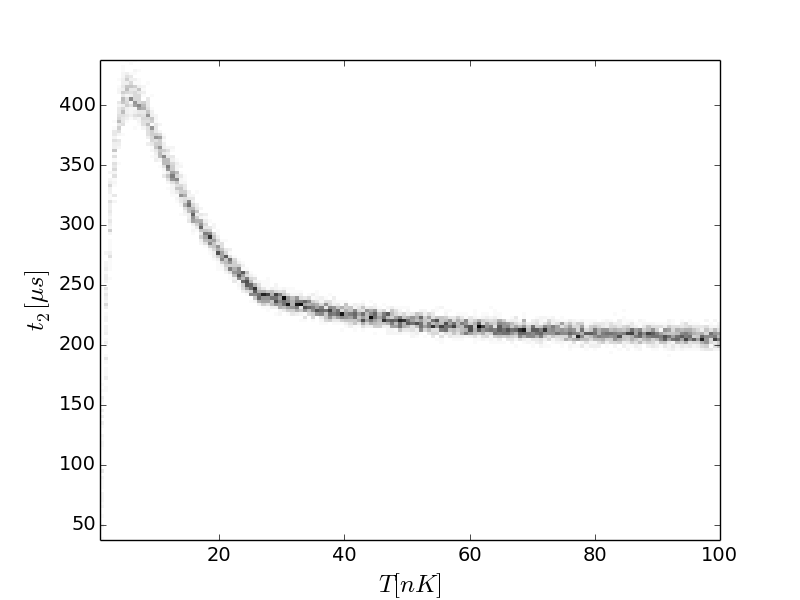}
\caption{(Color online) Figure shows the coherence phase $t_2 = \Gamma^{-1}_2$ as a function of temperature $T$ for a particle in the trap geometry $\omega_x,\omega_y = 2\pi\times42.0$ Hz and $\omega_z = 120.0$ Hz. 
For intermediate values of temperature, the coherence time is around $(300 - 450)~\mu$s (larger than the minimal time uncertainty of about $(50 - 100)~\mu$s). 
Above the critical temperature of about $T_c = 26.9$ nK, it approaches a value of less than $100~\mu$s converging to zero in the limit of very large temperatures.}
\label{figseven}
\end{center}
\end{figure}

Above the critical temperature, the real part of the fugacity spectrum shows a gap between the ring of constant absolute time, indicating that the symmetry of the quantum field can, in principle, not be broken without externally induced quantum fluctuations of energy and the corresponding entropy. Decreasing temperature by external cooling to decrease the gap between the outer ring and the inner linear (symmetry-breaking) part of the fugacity spectrum finally leads to a gapless fugacity spectrum. 
However, due to the uncertainty of the time variable, there is also a finite coupling probability between symmetric and asymmetric parts of the fugacity spectrum, in the parameter range, where the fugacity spectrum obeys a gap between the two principally unconnected spectra close above the critical temperature, which leads to spontaneous symmetry breaking, see Fig. \ref{figseven}. 

The probabilistic modeling of the dilute Bose gas with an average of $N$ particles below the critical temperature is described by the equation of conditional probabilities

\begin{equation}
\frac{p( \mu_1)}{p( \mu_2)} = \frac{{\rm e}^{-\beta\vert\mu_1\vert}}{{\rm e}^{-\beta\vert\mu_2\vert}} = \frac{{\rm e}^{t_1}}{{\rm e}^{t_2}} , 
\label{time_transition}
\end{equation}\\
in the present numerical framework, which defines the relative probability of a quantum particle to switch from a state $1$ with a corresponding chemical potential $\mu_1$ to a next state $2$ with a chemical potential $\mu_2$ for a Bose gas at thermal equilibrium, where the chemical potential is defined 
by the intrinsic particle number conserving equation for the average number of particles.

Applying only the constraint of particle number conservation as defined by the conservation equation in Eq. (\ref{eq.9}), numerical sampling leads to a distribution of typical scales for $\Gamma_1 t$ (oscillation time) and $\Gamma_2 t$ (coherence time), shown in Fig. \ref{figsix}. 
From the numerical sampling, it is observed that $\Gamma_1 t$ is proportional to the real part of the quantum phase and distributed around zero measure in the given parameter range (as defined in Fig. \ref{figsix} - upper figure). 
The real part $\Gamma_1 t$ defines the average oscillation frequency, which means that the quantum particle (with constant particle number at finite temperature) quickly couples to excited single particle quantum states, related to $\Gamma_1 t\sim0$ - in the form of a standing wave. 
The distribution of the phase $\Gamma_2 t$, that is, the effective coherence time for the particle weighted with the decay constant $\Gamma_2$, is shown in the lower figure of Fig. \ref{figsix}. 
The typical range of the time scale $\Gamma^{-1}_2$ is between $300 ~\mu s$ and $450 ~\mu s$ (see Figs. \ref{figsix} and \ref{figseven}) in the present setup. 
To estimate the time scale in a physical framework, time is scaled in terms of the quantity $\tau=2\pi\times(\omega_x\omega_y\omega_z)^{-1/3}$, 
since any frequency of the equilibrated system below that value is effectively zero in the thermally stable quantum system, because of the finite energy uncertainty.
From the present model, it is possible to, in particular, conclude that the so-derived time variable of a bosonic quantum field can provide a fundamental quantization of time only on the scale of quantized time steps $\tau_0$ defined by different realizations of the Boltzmann equilibrium.
Time intervals smaller than the value of $\tau_0$ randomly vary from effectively zero to tens of microseconds at different realizations of the quantum field, which leads to a quasi-continuous numerical time variable without a fundamental unit time (i.e., varying time step sizes) for time steps smaller than the time scale $\tau_0$.
    
\section{Discussion}

It is the Schr\"{o}dinger equation that describes the unitary time evolution of a quantum system, which follows the dynamics in terms of an energy operator. 
General solutions of this equation can be built up from coherent interference of partial specific solutions of the same equation.
In terms of the eigenvalues and eigenstates of the Schr\"{o}dinger equation, any general quantum state can be expressed in terms of the so-defined basis states, and hence 
evolves in time in a standard forward direction.

The fact that the conjugate equation precisely describes the backward time evolution of the adjoint quantum states is called the time-reversal symmetry of the Schr\"{o}dinger equation.
Very weakly interacting Bose gases that can be modeled and described in terms of a non-interacting Bose gas model build a fundamental theoretical system to study quantum effects such as spontaneous symmetry breaking on a measurable scale. 
The corresponding eigenstates and eigenenergies are not only straightforwardly defined and derivable formally, but can also be calculated numerically in terms of three-dimensional harmonic oscillator states.
From the definition of the quantum field in Eq. (\ref{eq.7}), which is a spatially integrated representation of the quantum field in the complex number plane, it is possible to illustrate that Glauber coherent states are built 
the specific solutions of the Schr\"{o}dinger equation for an ideal Bose gas with the complex variables $\alpha = \vert\psi\vert{\rm e}^{i\phi}$ that define the corresponding symmetric gauge fields as a function of the phase variable $\phi$. 

In a local representation of the quantum field, in very generality and mathematical and formal terms, the local symmetry of the quantum system can formally only be broken from a formal process that either modifies the fundamental equation of state or the corresponding gauge fields, respectively, called gauge fixing \cite{ref-16}.
In non-interacting Bose gases, such formal projection can, therefore, not only be understood by picking out a certain phase variable by a spooky unphysical action, such as the definition of a mathematical variable by an external observer.
As shown in the sequel of the present theory, following the ansatz of a non-local order parameter, the occurrence of spontaneously broken gauge fields can be ideally explained in physical terms by assuming 
the convergence of the (independent) quantum fields to the Boltzmann equilibrium with a vanishing total chemical potential at constant particle number and temperature of the Bose gas.
Thus, the symmetry of the considered wave field is spontaneously broken not from defining a specific formal (zero) gauge of the wave field, but from the physical inset of quantum coherence at the thermal equilibrium, 
while the underlying total symmetry of the wave field in complex number representation remains preserved.
Finally, the one-directional propagation of time is a direct formal consequence of the symmetry breaking of the Schr\"{o}dinger equation from the assumption of a Boltzmann equilibrium.
    
\section{Conclusion}

In summary, the presented particle-number conserving quantum field theory explains the concept of broken gauge symmetry as a physical rather than a purely formal and mathematical process from coherent interference of 
cold atomic matter waves and the equilibration of the Bose gas to the Boltzmann equilibrium.
Within the present theory, it was shown that spontaneous symmetry breaking doesn't rely only on a formal definition of a specific gauge of the wave field, but that symmetry breaking with a certain zero gauge of the wave field 
and one-directional time propagation naturally follows from the implementation of physical constraints in terms of particle number conservation and the assumption of the Boltzmann equilibrium. 
This is because, below the critical temperature, the manifolds of condensate and non-condensate quantum fields interfere destructively as a consequence of the increasing coherence time 
between different single-particle states and define a relative reference frame for the relative symmetry breaking between condensate and non-condensate aggregate phases in the Bose gas.
    
It is particularly interesting to formulate the present quantum field description in pioneering future works on dynamical localization in other systems, such as neuronal networks, out of thermal equilibrium in a completely different parameter regime. 
   
\acknowledgments

The author acknowledges the financial support from IU Internationale Hochschule for the lecturer position at the university, which has, in particular, enabled the formulation and editing of the present theory on symmetry aspects of ideal Bose gases. 

\bibliographystyle{unsrt}
\bibliography{references}

\end{document}